\documentclass[prb, twocolumn,preprintnumbers,amsmath,amssymb ]{revtex4}

\usepackage{graphicx}
\usepackage{dcolumn}
\usepackage{bm}

\begin{document}

\input epsf.sty



\title{Implications of spin vortex scenario for 1/8-doped lanthanum cuprates}

\author{Boris V. Fine}

\affiliation{ Institute for Theoretical Physics,University of Heidelberg, Philosophenweg 19, 69120 Heidelberg, Germany }

\date{4 March, 2009}

\begin{abstract}
Superlattice of spin vortices has been proposed in an earlier article as an alternative to the stripe interpretation of spin modulations in 1/8-doped lanthanum cuprates. The present article addresses several additional characteristics of the spin vortex lattice, namely: (i) the nature of extended charge states; (ii) the position of neutron spin peaks as a function of doping; and (iii) the absence of higher order spin harmonics. All these characteristics afford straightforward connection to the experimental results produced by angle-resolved photoemission spectroscopy,
resistivity measurements, and neutron scattering. 
\end{abstract}
\pacs{74.72.-h, 74.81.-g, 74.72.Dn}


\maketitle

\narrowtext
\pagebreak

\begin{figure} \setlength{\unitlength}{0.1cm}
\begin{picture}(100, 127) 
{ 
\put(5, 119){{\Large (a)}}
\put(13, 67){ \epsfxsize= 2.3in \epsfbox{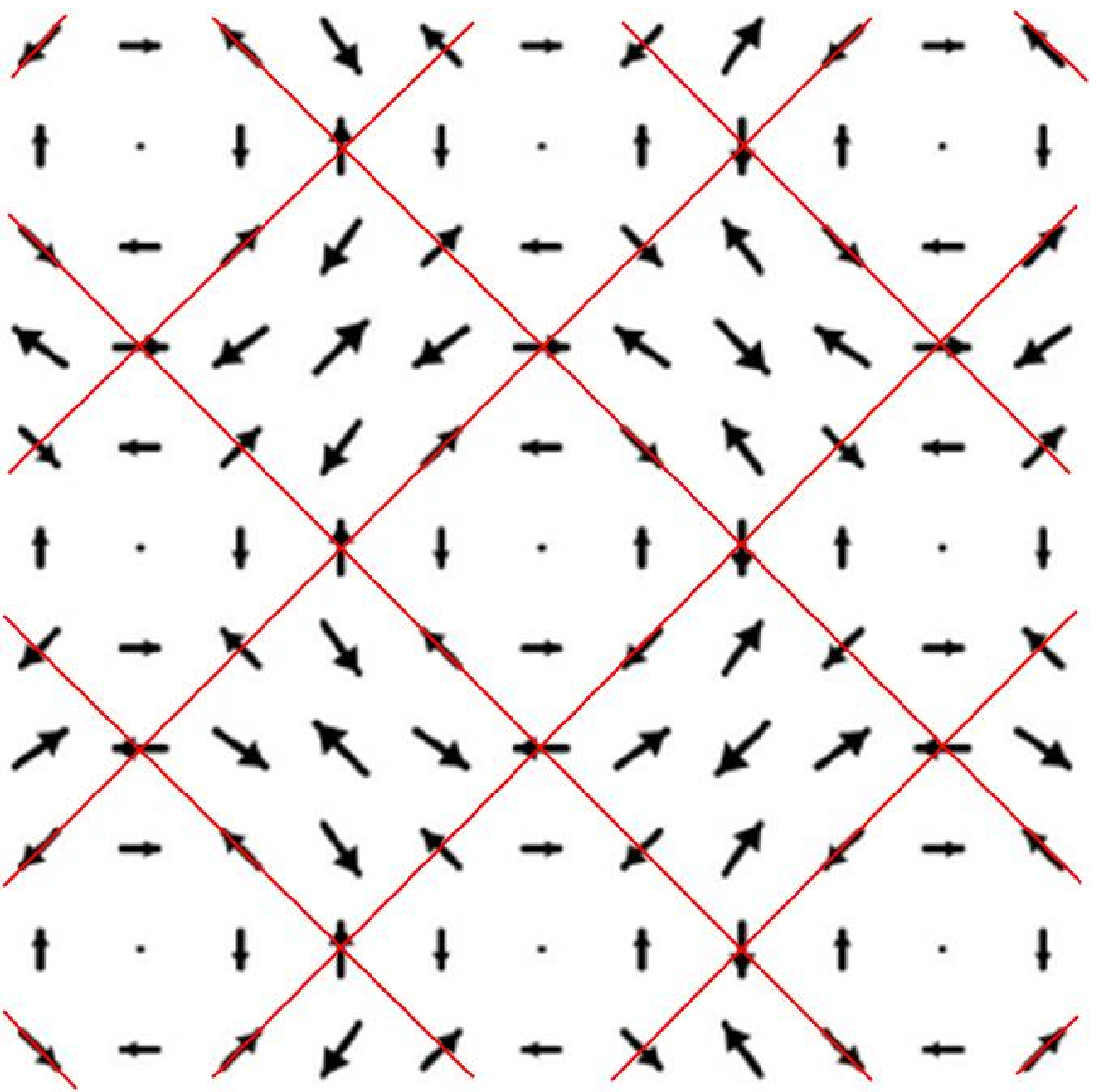  } }
\put(2, 51){{\Large (b)}}
\put(13, 0){ \epsfxsize= 2.3in \epsfbox{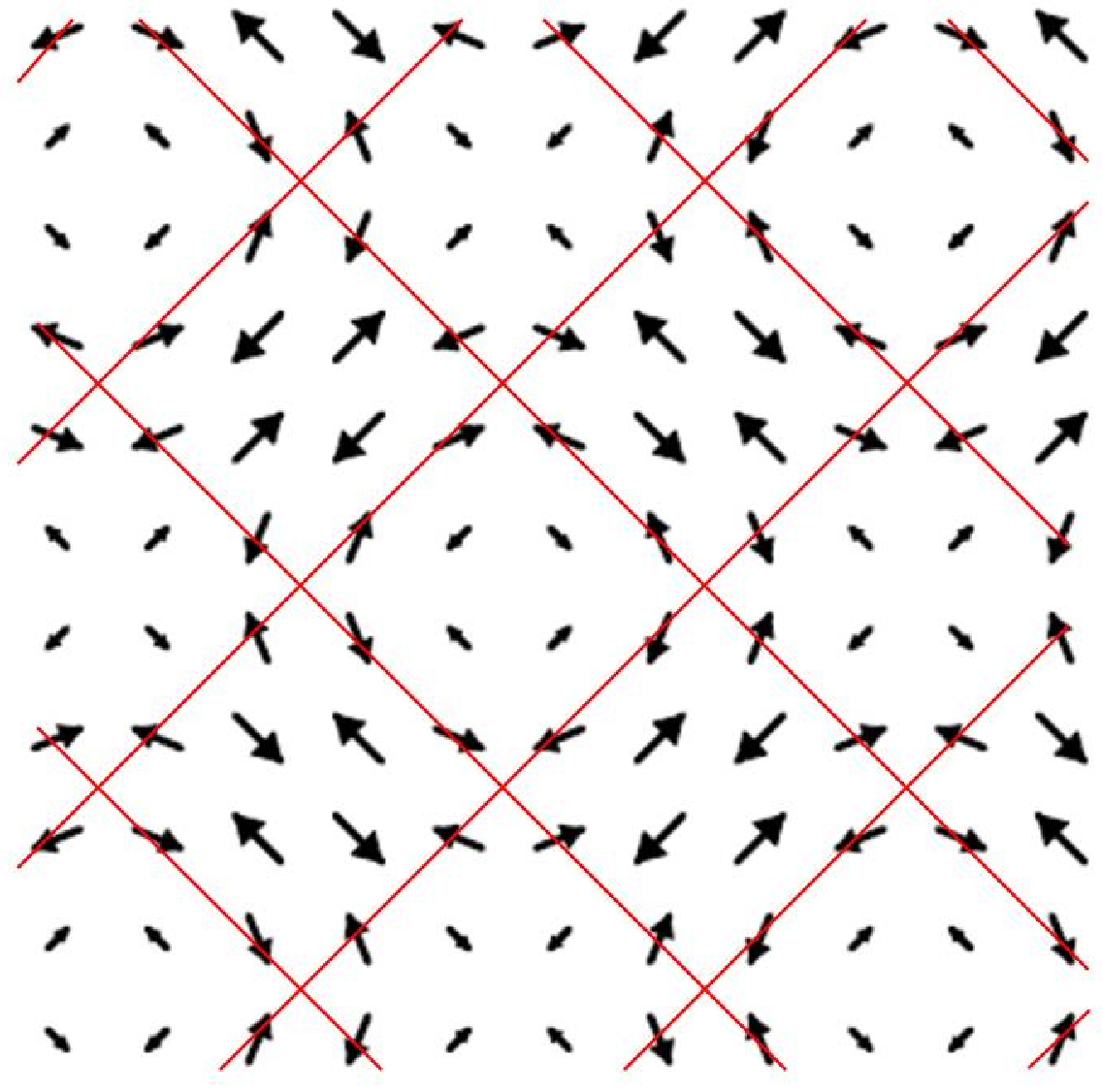  } }
}
\end{picture} 
\caption{(Color online) Two examples of commensurate spin vortex lattices: (a) Vortex cores site-centered. (b) Vortex cores plaquette-centered. Straight lines indicate the directions, where the structures exhibit perfect spirals.
} 
\label{fig-vortices} 
\end{figure}

The tendency of high-temperature cuprate superconductors towards the formation of electronic inhomogeneities is well known and discussed (see, e.g., Ref.\cite{Fine-Egami-08} and references therein). Yet the generic outcome of this tendency and its relation to the mechanism of superconductivity remains unclear. One can only be certain  that the implications of the inhomogeneous structures would be strongly dependent on their dimensionality. A group of materials that clearly show inhomogeneous static spin response in the bulk is 1/8-doped lanthanum cuprates. At present, the majority viewpoint is that this response originates from a pattern of one-dimensional spin and charge modulations called ``stripes''\cite{Tranquada-95}. This author had previously voiced doubt in the stripe interpretation in Refs.\cite{Fine-hitc-prb04,Fine-NQR-prb07}. The original consideration in the above references was based on a two-dimensional {\it spin-collinear} alternative to stripes called ``grid''. A later experiment has, however, produced a result incompatible with the grid superstructure\cite{Christensen-07}, but it did not rule out the possibility of a {\it non-collinear} two-dimensional superstructure.  In Ref.\cite{Fine-vortices-prb07}, this author has introduced a minimal realization of such a superstructure referred to below as ``spin vortex lattice'' (see Fig.~\ref{fig-vortices}), and further argued that it still constitutes a viable interpretation  of neutron scattering experiments in 1/8-doped lanthanum cuprates --- in particular, more straightforwardly than grid reproducing the positions of the charge peaks. [Reference \cite{Fine-vortices-prb07} refers to the same objects as ``magnetic vortices.'' Here I decided to change the terminology to avoid false associations with superconducting vortices.]

Further adding to the difficulties of the stripe interpretation, a very recent experiment\cite{Fujita-08} has reported static spin modulation in La$_{1.87}$Sr$_{0.13}$Cu$_{0.99}$Fe$_{0.01}$O$_4$ in the low temperature orthorhombic (LTO) phase rather than the low-temperature tetragonal (LTT) phase  --- the latter being the cornerstone in the system arguments supporting stripes\cite{Tranquada-95,Zimmermann-98}. The above finding is consistent with the view of the author\cite{Fine-vortices-prb07}, that it is not the geometry of the LTT phase as such but rather the proximity to the LT0-LTT phase transition that stabilizes the static spin response.

The purpose of this article is to communicate several additional facts and considerations related to the spin vortex scenario. The issues to be considered are: (i) Extended charge states; (ii) Position of neutron spin peaks as a function of doping; (iii) Absence of higher order spin harmonics.

The spin vortex lattice shown in Fig.~\ref{fig-vortices} is obtained for a square lattice of spins by a coherent superposition of two orthogonal spin harmonics with transverse in-plane polarizations:
\begin{eqnarray}
\mathbf S_{ij} = (-1)^{i+j} &
\left[
\; \;
\left(
\begin{array}{c}
0 \\
S_0
\end{array}
\right)
 \hbox{cos} \left( \mathbf q_1 \cdot \mathbf r_{ij} + \varphi_1 \right)
\right.
\nonumber
\\
&
\; \;
\left.
+
\left(
\begin{array}{c}
S_0 \\
0
\end{array}
\right)
\hbox{cos} \left( \mathbf q_2 \cdot \mathbf r_{ij} + \varphi_2 \right)
\right],
\label{S}
\end{eqnarray}
where $\mathbf q_1 = (q_0, 0)$ and $\mathbf q_2 = (0, q_0)$ are the wave vectors of the two harmonics with $q_0 = {\pi \over 4 a}$, $a$ is the lattice period, $S_0$ is the polarization amplitude of each harmonic, $\mathbf r_{ij}$ are the radius-vectors of the lattice sites with the origin chosen on the one of the sites, and $\varphi_1$, $\varphi_2$ are the phases.
Fig.~\ref{fig-vortices}a corresponds to $\varphi_1 = \varphi_2 = 0$ (site-centered vortices), Fig.~\ref{fig-vortices}b to $\varphi_1 = \varphi_2 = {1 \over 2} q_0 a$ (plaquette-centered vortices). Doping level 1/8 implies two doped holes per spin vortex.

This author has not attempted to derive the above structure from a microscopic treatment but rather looked at the experimental phenomenology. One can, however, learn about various theoretical considerations leading to the formation of spin vortices from Refs.\cite{Seibold-98,Berciu-99,Timm-00,Wilson-08,Wilson-08A,Koizumi-08,Koizumi-08A}. It remains a challenge though to understand how two holes would organize themselves within a single spin vortex.

{\it Extended charge states.} 

Spin vortex lattice commensurate with the underlying atomic lattice exhibits a network of diagonal lines, along which spins form perfect spirals --- see Fig.~\ref{fig-vortices}. These spiral lines can host gapless extended states of charge carriers and, therefore, explain the results of the recent angle-resolved photoemission spectroscopy (ARPES) study\cite{Valla-06}, which indicated that La$_{1.875}$Ba$_{0.125}$CuO$_4$  exhibits the usual pseudogap structure with gapless direction at 45 degrees with respect to the principal crystal axes. This ARPES result poses a serious challenge to the stripe picture.


The network of spirals can also explain why the resistivity of the sample drops on cooling below the temperature, where the system starts exhibiting static magnetic response\cite{Li-07}. Indeed, the static transition can be interpreted as the crystallization of vortices, which exist in the ``liquid'' form above the transition temperature. The liquid of spin vortices distorts the above network of spirals and, therefore, exhibits higher resistivity. The resistivity drop is then explained by a sudden formation of a perfect spiral network.

{\it Neutron peak splitting as a function of doping.} 

If one assumes, as normally done in the framework of the stripe interpretation, that the static spin superstructure observed at 1/8 doping also exists at other dopings in the fluctuating form and thus responsible for the inelastic neutron scattering peaks at $({\pi \over a}, {\pi \over a}  \pm {2 \pi \over a} \delta)$, $({\pi \over a}  \pm {2 \pi \over a} \delta, {\pi \over a})$, then the spin vortex picture leads to the dependence of the dimensionless splitting parameter $\delta$  on  the doping concentration $x$ (per in-plane Cu atom),  which is different from $\delta = x$ obtained for quater-filled stripes. 

\begin{figure} \setlength{\unitlength}{0.1cm}
\begin{picture}(100, 60) 
{ 
\put(0, 0){ \epsfxsize= 3.2in \epsfbox{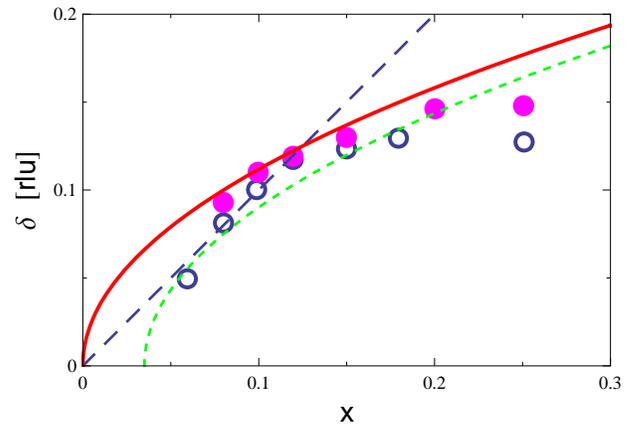  } }
}
\end{picture} 
\caption{
(Color online) Dependence of neutron peak splitting parameter $\delta$ on the doping concentration $x$. Filled circles represent the experimental results in La$_{1.6-x}$Nd$_{0.4}$Sr$_x$CuO$_4$\cite{Birgeneau-06}, empty circles --- in La$_{2-x}$Sr$_x$CuO$_4$\cite{Yamada-98}. Solid line represents Eq.(\ref{delta}),
small-dash line Eq.(\ref{delta1}) with $x_{c0}=0.35$, large-dash line is the linear dependence $\delta = x$.
}
\label{fig-yamada} 
\end{figure}

Spin vortex lattice contains two holes per vortex. If the vortex cores are spaced by distance $l$  (in units of $a$), and all doped holes are captured by the vortices, then $x = {2 \over l^2}$, and the magnetic periodicity is $2 l$, i.e. $\delta = {1 \over 2l}$. Therefore,
\begin{equation}
\delta = \sqrt{{1 \over 8} x}.
\label{delta}
\end{equation}
This dependence is compared with the experimental results for La$_{2-x}$Sr$_x$CuO$_4$\cite{Yamada-98} and La$_{1.6-x}$Nd$_{0.4}$Sr$_x$CuO$_4$\cite{Birgeneau-06} in Fig.~\ref{fig-yamada}. 

It is unrealistic to expect that all  doped holes are captured inside spin-poor regions of spin superstructures. Some may remain delocalized in spin-rich regions. In this case, the above expressions would overestimate $\delta$ both for stripes and for spin vortices. Therefore, the $\delta$-vs.-$x$ scaling as such cannot conlusively confirm or rule out any of the two propositions. Yet, the experimental data  shown in Fig.~\ref{fig-yamada} appear to lend somewhat stronger support to the spin vortex scaling.


One can further assume that only holes added in excess of certain concentration $x_0$ are captured by the vortices. In this case
\begin{equation}
\delta = \sqrt{{1 \over 8} (x-x_0)}.
\label{delta1}
\end{equation}
An example of such dependence with $x_0 = 0.035$ is plotted in Fig.2. Concentration $x_0$ itself can and should further depend on the net doping level $x$. However, it is not unreasonable to assume that it stays approximately the same in not too broad range of $x$, e.g. between $x = 0.1$ and $x = 0.15$.

{\it Absence of higher order spin harmonics.} 

Another outstanding issue that needs to be understood within the frameworks of both stripes and spin vortex interpretations is the lack of evidence for the higher order spin harmonics\cite{Tranquada-98}. In general, the higher order peaks have intensity smaller than the main ones, but given the strength and the sharpness of the latter, it is difficult to understand the complete lack of evidence for the former. Higher order harmonics have been observed for stripe-ordered nickelates\cite{Wochner-98}.

Spin vortex lattice does offer a good argument against the presence of higher order spin harmonics: As explained below, one cannot sustain a non-collinear pattern of staggered spin polarizations, if only a few harmonics are added to the main ones. One needs a simultaneous contribution of many additional harmonics of comparable amplitude, in order to satisfy a local stability condition. In such a case, the harmonics with the lowest susceptibility (highest energy cost) would control the higher order spin response, and therefore, this response would be anomalously low.

I assume the nearest-neighbor exchange interaction characterized by energy
\begin{equation}
E = J \sum^{_{\hbox{\small NN}}}_{ij,m>i,n>j} \mathbf S_{ij} \cdot \mathbf S_{mn}
\label{E}
\end{equation}
where $(ij)$ and $(mn)$ are the pairs of square lattice indices, $\mathbf S_{ij}$ are the local staggered spin polarizations, and $J$ is the exchange constant. Superscript ``NN'' indicates that sites $(mn)$ are the nearest neighbors of sites $(ij)$. 
Given Eq.(\ref{E}), each spin experiences local field:
\begin{equation}
\mathbf h_{ij} = J \sum^{_{\hbox{\small NN}}}_{mn} \mathbf S_{mn}.
\label{h}
\end{equation}

A non-collinear spin configuration is stable locally, when $\mathbf h_{ij} \parallel \mathbf S_{ij}$ or
\begin{equation}
 S_{ij,x} h_{ij,y} - S_{ij,y} h_{ij,x} = 0,
\label{Sh}
\end{equation}
where subscripts $x$ and $y$ indicate the projections of respective vectors.
If this condition is violated, the local polarization cannot  be static, because it would start precessing  around the local field direction.  

It has been shown in Ref.\cite{Fine-vortices-prb07} that the above condition is satisfied for a superposition of any two non-helical sinusoidal harmonics. Here I consider what happens in the case of an arbitrary superposition of helical or non-helical harmonics with in-plane spin polarization having form
\begin{eqnarray}
\mathbf S_{ij} &=& (-1)^{i+j} \sum_{\alpha} 
\left[
\begin{array}{c}
S_{\alpha x}  \hbox{cos} \left( \mathbf q_{\alpha} \cdot \mathbf r_{ij} + \varphi_{\alpha x} \right)
\\
S_{\alpha y}  \hbox{cos} \left( \mathbf q_{\alpha} \cdot \mathbf r_{ij} + \varphi_{\alpha y} \right)
\end{array}
\right]
\nonumber
\\
&
\equiv 
&
\sum_{\alpha} \mathbf S_{\alpha} (\mathbf r_{ij}),
\label{Salpha}
\end{eqnarray}
where $\alpha$ is the index of spin harmonics, $\mathbf q_{\alpha}$ the wave vectors; $S_{\alpha x}$, $S_{\alpha y}$ and $\varphi_{\alpha x} $, $\varphi_{\alpha y} $ are, respectively, the amplitudes and the phases for the $x$- and $y$-components of a given harmonic; $\mathbf S_{\alpha} (\mathbf r_{ij})$ denotes the entire spin field  of one harmonic.

According to Eq.(\ref{h}), the local field corresponding to the static spin pattern given by Eq.(\ref{Salpha}) is
\begin{equation}
\mathbf h_{ij} = -2 J \sum_{\alpha} 
\left[
\hbox{cos} 
\left( 
\mathbf q_{\alpha} \cdot \mathbf a
\right)  
+ \hbox{cos} 
\left( 
\mathbf q_{\alpha} \cdot \mathbf b
\right)  
\right]
\mathbf S_{\alpha} (\mathbf r_{ij}), 
\label{halpha}
\end{equation}
where $\mathbf a$ and $\mathbf b$ are the primary translation vectors along the two principal directions of the underlying square lattice. [Note: If the exchange with more remote neighbors were included in energy (\ref{E}), it would add terms such as, e.g. 
$\hbox{cos} 
\left( 
\mathbf q_{\alpha} \cdot (\mathbf a + \mathbf b)
\right)  $
or  
$\hbox{cos} 
\left( 
\mathbf q_{\alpha} \cdot 2 \mathbf a
\right)  $
to the prefactor in the square bracket in Eq.(\ref{halpha}). The presence of these terms would not change the conclusions that follow.]

The substitution of expressions (\ref{Salpha}) and (\ref{halpha}) into the local stability condition (\ref{Sh}) gives 
\begin{widetext}
\begin{eqnarray}
\sum_{\alpha, \beta > \alpha}
\left[
\hbox{cos} 
\left( 
\mathbf q_{\beta} \cdot \mathbf a
\right)  
+ \hbox{cos} 
\left( 
\mathbf q_{\beta} \cdot \mathbf b
\right)
-
\hbox{cos} 
\left( 
\mathbf q_{\alpha} \cdot \mathbf a
\right) 
- \hbox{cos} 
\left( 
\mathbf q_{\alpha} \cdot \mathbf b
\right)  
\right]
\;\;\;\;\;\;\;\;\;\;\;\;\;\;\;\;\;\;\;\;\;\;\;\;\;\;\;\;\;\;\;\;\;\;\;\;\;\;\;\;
\;\;\;\;\;\;\;\;\;\;\;\;\;\;\;\;\;\;\;\;\;\;\;\;\;\;\;\;\;
&&
\nonumber
\\
 \;\;\;\;\;\;\;\;\;\;
\times
\left[
S_{\alpha x} S_{\beta y} 
\hbox{cos} \left( \mathbf q_{\alpha} \cdot \mathbf r_{ij} + \varphi_{\alpha x} \right)
\hbox{cos} \left( \mathbf q_{\beta} \cdot \mathbf r_{ij} + \varphi_{\beta y} \right)
-
S_{\alpha y} S_{\beta x} 
\hbox{cos} \left( \mathbf q_{\alpha} \cdot \mathbf r_{ij} + \varphi_{\alpha y} \right)
\hbox{cos} \left( \mathbf q_{\beta} \cdot \mathbf r_{ij} + \varphi_{\beta x} \right)
\right]
&=& 0.
\label{Sum}
\end{eqnarray}
\end{widetext}

There are two particularly simple situations, when condition (\ref{Sum}) is satisfied. The first one corresponds to all harmonics collinear along the same direction. In this case the second square bracket in each term of Eq.(\ref{Sum}) equals zero. Both stripe and grid interpretations fulfill this condition, and, therefore, the local stability does not prevent either stripes or grid from having any combination of higher order harmonics collinear with the main ones. 

The second simple situation corresponds to the first square bracket in Eq.(\ref{Sum}) being equal to zero for all pairs of $\alpha$ and $\beta$. This condition is satisfied, for example, for any two harmonics commensurate or incommensurate, helical or not, provided $\mathbf q_{\alpha}$ and $\mathbf q_{\beta}$ are rotated with respect to each other by 90 degrees, and $|\mathbf q_{\alpha}| = |\mathbf q_{\beta}|$. Therefore, any non-collinear structure involving only two suchharmonics $\mathbf q_1$ and $\mathbf q_2$ (given earlier) is locally stable. 

The wave vectors of potential higher order harmonics have form
\begin{equation}
\mathbf q_{\alpha} = \left( \pm {2 \pi \over 8 a} n, \pm {2 \pi \over 8 a} m  \right); \ \ \ \hbox{with} \ \ n,m = 0,1,2,3,4 ;
\label{qalpha}
\end{equation}
except for the pairs $(n=\pm 1, m=0)$ and $(n=0, m=\pm 1)$  corresponding to the wave vectors $\pm \mathbf q_1$ and $\mathbf q_2$, respectively.

When two non-collinear harmonics $\mathbf q_1$ and $\mathbf q_2$ are already present, none of the higher order harmonics (\ref{qalpha}) would make the first square bracket in Eq.(\ref{Sum}) equal to zero for the term involving this higher order harmonic and either $\mathbf q_1$ or $\mathbf q_2$. 
In this case, condition (\ref{Sum}) cannot be satisfied on term-by-term basis for every pair of harmonics $\alpha$ and $\beta$. However, it is still possible to make the total sum (\ref{Sum}) identically zero for every value of $\mathbf r_{ij}$. The Fourier transform of $8 \times 8$ spin superstructure has maximum 64 wave vectors. When all possible harmonics contribute to the spin superstructure, there should exist infinitely many ways of fulfilling condition (\ref{Sum}), because two polarization directions of the contributing harmonics ($x$ and $y$) imply 128 adjustable parameters (amplitudes and phases) to cancel 64 Fourier components of the sum (\ref{Sum}). The non-collinear superstructure considered in Ref.\cite{Christensen-07} consisting of disconnected collinear magnetic clusters, is one such a locally stable example of adding higher order harmonics to the ones at $\mathbf q_1$ and $\mathbf q_2$.
There, in fact, every cluster can have an arbitrary polarization with respect to the others and still remain locally stable. Such a superstructure does contain sizable contributions from many higher order spin harmonics, none of which has been observed experimentally.

It is now important to realize, that one cannot simply add a small number of higher order harmonics to the two at $\mathbf q_1$ and $\mathbf q_2$ and still preserve the local stability. The reason is the following:

Each $(\alpha,\beta)$-term in sum (\ref{Sum}) contains a product of $\hbox{cos} ( \mathbf q_{\alpha} \cdot \mathbf r_{ij} + \varphi_{...})$ and $\hbox{cos} ( \mathbf q_{\beta} \cdot \mathbf r_{ij} + \varphi_{...})$ and, therefore, contributes to the Fourier components of the sum at $q_{\alpha} + q_{\beta}$ and $q_{\alpha} - q_{\beta}$. The contributions to a given Fourier component can thus come from different pairs $(\alpha,\beta)$, which need to cancel each other. Therefore, there should be either zero or at least two such contributions to a given Fourier component.  With too few harmonics added to $\mathbf q_1$ and $\mathbf q_2$, there will necessarily be values of $q_{\alpha} \pm q_{\beta}$, where only one  $(\alpha,\beta)$-term contributes, and therefore the local stability condition cannot be satisfied. 

I was not able to establish rigorously the minimum number of higher order harmonics, to cancel only the leading contributions to sum (\ref{Sum}) from pairs involving $\mathbf q_1$ or $\mathbf q_2$, but from experimenting with various possibilities, the set harmonics located at 
\begin{equation}
\mathbf q_{\alpha} = \left( \pm {2 \pi \over 4 a} n, \pm {2 \pi \over 4 a} m  \right); \ \ \ \hbox{with} \ \ n,m = 0,1,2 ;
\label{qalpha2}
\end{equation}
appears to be the  most economic proposition. It is clear in any case, that one needs to introduce a large number of harmonics of comparable amplitude spanning the entire Brillouin zone, which brings us back to the argument made earlier that the susceptibility to such a configuration should be anomalously low.

The author is grateful to E. Carlson, T. Egami, M. Fujita, H. Koizumi and G. Seibold for discussions and communications related to this work.

\bibliography{vor2}

\end{document}